\documentclass[aps,prb,twocolumn,superscriptaddress,showpacs]{revtex4}
\usepackage{graphicx}

\begin{document}

\title{Orbital magnetic properties of quantum dots: the role of electron-electron interactions}

\author{L. G.~G.~V. Dias da Silva}
\affiliation{Departamento de F\'{\i}sica, Universidade Federal de
S\~ao Carlos, 13565-905 S\~ao Carlos SP, Brazil}

\author{Caio H. Lewenkopf}
\affiliation{Instituto de F\'{\i}sica, Universidade do Estado do
Rio de Janeiro, R. S\~ao Francisco Xavier 524, 20550-900 Rio de
Janeiro, Brazil}

\author{Nelson Studart}
\affiliation{Departamento de F\'{\i}sica, Universidade Federal de
S\~ao Carlos, 13565-905 S\~ao Carlos SP, Brazil}

\date{\today}

\begin{abstract}
We study the magnetic orbital response of a system of $N$
interacting electrons confined in a two-dimensional geometry and
subjected to a perpendicular magnetic field in the finite
temperature Hartree-Fock approximation. The electron-electron
interaction is modelled by a short-range Yukawa-type potential. We
calculate the ground state energy, magnetization, and magnetic
susceptibility as a function of the temperature, the potential
range, and the magnetic field. We show that the amplitude and
period of oscillations in the magnetic susceptibility are strongly
affected by the electron-electron interaction as evidenced in
experimental results. The zero-field susceptibility displays both
paramagnetic and diamagnetic phases as a function of temperature
and the number of confined electrons.
\end{abstract}



\maketitle

\newcommand{\be}   {\begin{equation}}
\newcommand{\ee}   {\end{equation}}
\newcommand{\ba}   {\begin{eqnarray}}
\newcommand{\ea}   {\end{eqnarray}}
\newcommand{\HF}   {\mbox{\scriptsize HF}}

\section{Introduction}
\label{sec:introduction}

Much effort is currently devoted to the study of electron-electron
interaction effects on the ground state and transport properties
of mesoscopic and nanostructured devices.
\cite{Mesoreview,Aleiner02,Ullmo01} The theoretical analysis of
small quantum dots, with $N \alt 8$ electrons, costumarily employs
sophisticated numerical tools. \cite{Reimann02} Those become
computationally prohibitive for larger dots, that call for a more
schematic approach, such as mean-field and/or semiclassical
approximations. In this paper we address the specific question of
how the electron-electron interactions change the magnetic
susceptibility of quantum dots using the self-consistent
Hartree-Fock approximation.

Our motivation stems mainly from the intriguing experimental data
collected in the early 90's by L\'evy and collaborators,
\cite{Levy93} who measured the orbital magnetic susceptibility of
an array of mesoscopic squares lithografically inscribed in a
GaAs/AlGaAs heterostructure. The authors found a surprisingly
large paramagnetic susceptibility and a power law dependence of
the zero-field susceptibility $\chi_0(B) \equiv \chi(B=0,T)$ with
temperature. The first result is understood by means of a
semiclassical single-particle analysis, showing that the magnetic
susceptibility in ballistic devices is determined by the enclosed
area and the stability of the shortest classical periodic orbits
of the cavity.
\cite{Prado94,vonOppen94,Agam94,Ullmo95,Gurevich97,Terra98} More
specifically, the cavity geometry determines the classical
periodic orbits relevant for $\chi(B,T)$, provided $B$ is weak.
For generic integrable systems such orbits come in families and
the magnitude $\chi(B,T)$ scales as $k_F L$, where $k_F$ is the
wave number at the Fermi energy and $L$ is the typical length
scale of the cavity. In contrast, periodic orbits of chaotic
systems are isolated and $\chi(B,T) \propto (k_FL)^{1/2}$. The
semiclassical theoretical analysis also predicts an exponential
decay for $\chi_0(T)$, which conflicts with the experimental data
that display a much weaker temperature dependence. The
experimental results \cite{Levy93} raised natural and important
questions, concerning the role of disorder and interactions. These
issues triggered interesting advances in the semiclassical
approach. The disorder in the devices we are interested in is weak
and long ranged, and hence dominated by small angle scattering. It
was convincingly shown \cite{Richter96b} that such kind of
disorder has little influence on the semiclassical results derived
for perfectly clean systems. A more far reaching and fundamental
question is the role electron-electron interactions. In a
sophisticated semiclassical approximation to the many-body
problem, Ullmo and collaborators \cite{Ullmo98} showed that
interactions make $\chi(B,T)$ scale as $k_F L$ for integrable
systems and $k_F L/\ln(k_F L)$ for chaotic ones. There two key
approximations that allow for a full analytical treatment
presented in Ref. \onlinecite{Ullmo98}. One approximation is to
consider a zero-range residual interaction, that beclouds the
exchange interaction, and the other is to give up self
consistency.

Further motivation stems from a more recent experiment on the
orbital magnetization of quantum-dot arrays \cite{Grundler02} that
found a magnetization value two orders of magnitude larger than
the one predicted by the noninteracting single-particle picture.
Existing theoretical predictions based on the direct numerical
diagonalization of the Hamiltonian for a few electrons
\cite{Maksym92,Wagner92,Creffield00,Gregorio02} and on a mean
field approximation of the many-body problem \cite{Fogler94}
indicated that the electron-electron interaction plays dominant
role in the magnetic properties of quantum dots as clearly
observed in the experiment. \cite{Grundler02}

In this work, we resume the discussion on the influence of
electron-electron interactions by studying the orbital magnetic
response of a $N$-electron interacting system in a self-consistent
Hartree-Fock (SCHF) approximation. The screened Coulomb
electron-electron interaction is modelled by the short-range
Yukawa potential $V(r)=V_0 e^{-\kappa r}/r$. Exact results are
only known for the $N=2$ case. \cite{Creffield00,Gregorio02} The
mean field approximation allows us to study systems up to $N \sim
40$ interacting electrons. Although the typical experimental dots
in Refs. [\onlinecite{Levy93,Grundler02}] have a much larger
number of electrons, our study provides a qualitative
understanding of the interaction effects in the magnetic
properties of the system.

We calculate the ground state energy $E_g$, magnetization, and
magnetic susceptibility as a function of the relevant parameters
of the system, namely, the temperature $T$, the potential range
$\kappa$ and the magnetic field $B$. Our results show
discontinuities in $E_g(B)$ at $T=0$, as previously reported.
\cite{Ahn99} These features are smoothed out for finite
temperature.

We show that the magnetic response of a two-dimensional electronic
cavity depends strongly on the electron-electron interaction and
the magnetic susceptibility shows an oscillatory behavior similar
to de Haas-von Alphen oscillations in metals.
Furthermore, the amplitude and period of such oscillations are
modified by the electron-electron interaction.
We discriminate the kinetic, direct, and exchange contributions to
the total magnetic susceptibility.
The direct and exchange contributions also oscillate but with a
different phase compared to the kinetic contribution. We also find
that the susceptibility dependence on temperature and interaction
strength displays non-universal features which are strongly
dependent on the parameters. A slight enhancement of the
zero-field susceptibility is seen as the number of electrons
increases. The interaction-induced magnetic susceptibility shows
paramagnetic and diamagnetic phases both as a function of
temperature and interaction strength. A rather unexpected result
is the behavior of the exchange interaction contribution to
$\chi(B)$, which becomes larger than the direct contribution as
the interaction strength increases.

The paper is organized as follows. In Sec.\ \ref{sec:meanfield} we
present the model system considered in this study and the mean field
solution in the SCHF approximation. The main results are shown in
Sec. III. Section IV brings the final remarks and conclusions. We
also include an appendix where we present some specific details of
the SCHF numerical implementation in the presence of discrete
symmetries.

\section{The model}
\label{sec:meanfield}

We consider the problem of $N$ two-dimensional (2D) interacting
electrons in a confining potential subjected to an  external
magnetic field ${\bf B}$ perpendicular to the electron system.
Since we study the orbital contribution to the magnetization, we
are allowed to simplify the problem and treat the electrons as
spinless. The model Hamiltonian reads as
\be
H = \sum_{n=1}^N h({\bf  r}_n) +
          \sum_{n < {n^\prime}}^N v({\bf r}_n, {\bf r}_{n^\prime})\,,
\ee where ${\bf r}_n$ indicates the position of the $n$th
electron. The single-particle Hamiltonian $h$ is given by
\be
h({\bf r}) = \frac{1}{2m^*}\left[{\bf p} +
        \frac{e}{c}{\bf A}({\bf r})\right]^2 + u({\bf r}) \,,
\ee
where $m^*$ is the electron effective mass. The vector potential
${\bf A}$ is chosen in the symmetric gauge, namely, ${\bf
A}=(-By/2,Bx/2,0)$. The magnetic field is expressed in units of
$\Phi/\Phi_0$, where $\Phi=B {\cal A}$ is the magnetic flux
through the system area ${\cal A}$ and $\Phi_0 = hc/e$ is
the unit quantum flux. We choose the confining potential $u({\bf
r})$ as the 2D square well of side $L$ that closely models the
experiment. \cite{Levy93}

To account for screening effects the electron-electron interaction
$v$ is modelled by \cite{Gregorio02}
\be
\label{eq:defv}
\label{eq:interacao}
v({\bf r},{\bf r}^\prime) = \frac{e^2}{4 \pi \epsilon_0
\epsilon_r} \frac{e^{ - \kappa |{\bf r}-{\bf r}^\prime|}} {|{\bf
r}-{\bf r}^\prime|} \,,
\ee
where $\kappa$ gives the effective interaction range and
$\epsilon_r$ is the background dielectric constant. For $\kappa =
0$ there is no screening and the bare Coulomb interaction is
recovered. Even though the $v({\bf r},{\bf r}^\prime)$ in
Eq.~(\ref{eq:defv}) is different from the screened
electron-electron interaction in the 2DEG when effects of the
finite layer-thickness and image charges are taken into account,
\cite{Ando82} the Yukawa potential captures the main features of
a more realistic interaction and has the advantage of being
computationally more amenable to handle. \cite{Candido98} We
recall that the semiclassical approach is forced to use zero-range
residual interaction in order to make the calculations feasible.
By doing so, one looses a handle on the exchange interaction.

For a square dot of side $L$, the potential energy scales with
$1/L$ while the kinetic energy scales with $1/L^2$. Hence, as $L$
is increased, the potential energy becomes increasingly more
important. It is then useful to introduce \cite{Ahn99} an
``effective strength" parameter $L/a^{*}_B$, where $a^{*}_B =
\hbar^2 (4 \pi \epsilon_0 \epsilon_r) / {m^* e^2}$ is the
effective Bohr radius. The standard dimensionless parameter
that quantifies the ratio between electronic potential and kinetic
energies is $r_s$, that in 2D reads $r_s^2={\cal A}/(N \pi[a^{*}_B]^2)$.
Hence, $L/a^{*}_B$ and $r_s$ are related as $r_s=(L/a^*_B)/
\sqrt{\pi N}$. We consider here a range of
parameters such that $1.5<r_s<2$, within the typical values of $r_s$
in the experiments. \cite{Levy93,Grundler02}

We calculate the ground-state energy in the SCHF approximation for
finite temperatures. The SCHF equations read as
\cite{Tamura97,Dean01}
\begin{eqnarray}
h({\bf r}) \phi_i({\bf r}) + \sum_j \left[ n_j \!
   \int d{\bf r}^\prime \phi_j^*({\bf r}^\prime)v( {\bf r},{\bf r}^\prime)
   \phi_j({\bf r}^\prime)\right]\phi_i({\bf r}) \nonumber \\
 - \sum_j \left[ n_j\! \int\!\! d{\bf r}^\prime \phi_j^*({\bf r}^\prime)
   v({\bf r},{\bf r}^\prime)\phi_j({\bf r})\phi_i({\bf r}^\prime) \right]
 = \varepsilon_i^{\HF} \phi_i({\bf r}) ,
\label{eq:HFeq}
\end{eqnarray}
where the sums run over all HF orbitals. Here $n_i =
\{\exp[(\varepsilon^{\rm HF}_i - \mu)/k_BT] + 1\}^{-1}$ is the
Fermi occupation number of the $i$th HF orbital with corresponding
wave function $\phi_i({\bf r})$ and energy $\varepsilon^{\HF}_i$.
As standard, the chemical potential $\mu$ is determined by
requiring that $N = \sum_i n_i$. We truncate the number of
orbitals and take only the $M \approx 2N$ lowest energy states
into account.

The SCHF ground state energy is given by
\ba
E^{\HF}_g & \equiv & T^{\HF} + V^{\HF}_d - V^{\HF}_x \nonumber\\
       & = & \sum_{i} n_i \langle\phi_i| h |\phi_i\rangle
             + \frac{1}{2} \sum_{i,j} n_i n_j \Big(
       \langle\phi_i \phi_j| v |\phi_i \phi_j\rangle \nonumber\\
       & & - \langle\phi_i \phi_j| v |\phi_j \phi_i\rangle \Big),
\label{eq:EHFdireto}
\ea
where the $|\phi_i\rangle$ are the HF orbitals, self-consistent
solutions of Eq.\ (\ref{eq:HFeq}) and $T^{\HF}$, $V^{\HF}_d$ and
$V^{\HF}_x$ are the kinetic, direct and exchange contributions to
the ground state energy respectively. Actually, it is numerically
less expensive to compute $E^{\HF}_g$ as
\be
E^{\HF}_g = \frac{1}{2} \sum_{i} n_i \Big( \varepsilon_i^{\HF} +
                      \langle\phi_i| h |\phi_i\rangle \Big) \,.
\label{eq:EHFcompac} \ee

The calculations of the matrix elements were done numerically,
with no further approximations other than setting the numerical
precision. The number of elements grows as $M^4$,
where $M$ is the basis size. For a typical calculation where we
take $M \sim 50$, a numerical evaluation of about $10^{7}$
four-dimensional integrals is required, a task far from trivial.
To reduce that number, we have used an appropriate symmetrized
basis on which the HF potential takes a block-diagonal form.
Furthermore, we use properties of the the Yukawa-like e-e interaction
to reduce the four-dimensional integrals to a series of one-dimensional
integrals on the relative polar angle.
These steps are described in Ref. \onlinecite{Gregorio02}.
Details on the HF numerical implementation in the presence of
discrete symmetries are found in the Appendix.

\section{Single-particle properties}
\label{sec:singleparticleproperties}

The HF single-particle spectral properties play a major role for
determining the system magnetization and magnetic susceptibility.
In this section we present a general discussion of the model
Hamiltonian single-particle spectrum that serves as a guide to
interpret the magnetic results that follow.

\begin{figure}[h!]
\includegraphics[width=8cm]{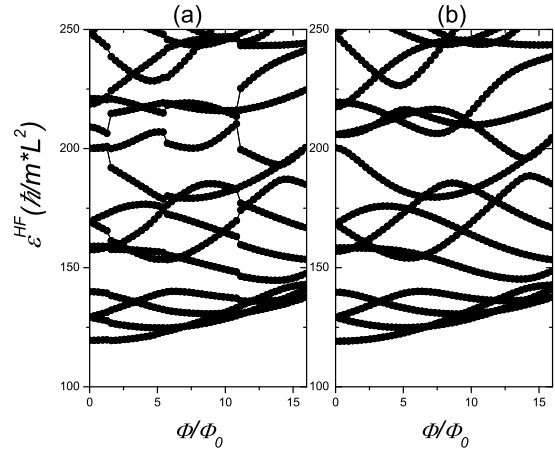}
\caption{ Single-particle Hartree-Fock energy levels
$\varepsilon^{\rm HF}_i$ as a function of the magnetic
flux $\Phi/\Phi_0$ for (a) $T=0$ and (b) $k_BT = \Delta$.
Here $N=10$ and $r_s=1.22$.}
\label{fig:EspHF}
\end{figure}

Figure \ref{fig:EspHF} shows the single-particle HF spectrum as a
function of the magnetic flux $\Phi/\Phi_0$ for $N=10$ electrons with
$r_s=1.22$ for $T=0$ (left) and $T = \Delta$ (right). Here
and throughout the manuscript $\Delta$ is the single-particle mean
level spacing. Energy is given in units of $\hbar^2/(m^* L^2)$.

A very interesting feature displayed in Fig.~\ref{fig:EspHF} are
the jumps in $\varepsilon^{\rm HF}_i(B)$ for the $T=0$ case. These
discontinuities appear for other values of $N$ as well and are
related to sudden anti-crossings at Fermi energy, {\it
i.e.}, involving the levels $\varepsilon^{\rm HF}_N$ and
$\varepsilon^{\rm HF}_{N+1}$.
For $T=0$ the mean field selects the $N$ lowest states. Hence, by
parametrically moving through a level crossing, the character of
the last occupied state is suddenly changed, and so is the mean-field.
Narrow crossings, with a gap $\delta \varepsilon \ll \Delta $,
that lead to sudden and very strong changes in the last occupied
mean-field state, are responsible for the jumps.

The jumps disappear already for very low temperatures of the order
of $k_BT \approx\delta \varepsilon \ll \Delta$, as illustrated in
Fig.\ \ref{fig:EspHF}(b). For this reason, they are of very limited
relevance for the experiments we are interested in, where $k_BT \gg
\Delta$.
However, Fermi energy anti-crossings explain similar features
reported in the study of ground-state properties of quantum dots
in the Coulomb blockade regime, that remained so far not understood.
\cite{Ahn99} Albeit also blurred
by temperature, the later case deals with transport where the
ground state many-body wave function matters and thus the Anderson
orthogonality catastrophe can come into play.
\cite{Vallejos02,Gefen02}

\section{Ground-state properties}
\label{sec:groundstateconf}

The magnetization $m(B)$ and the magnetic susceptibility
$\chi(B)$ of an electronic cavity are obtained from
\be
\label{eq:defchi}
m(B)  = -\frac{\partial \Omega}{\partial B} \qquad \mbox{and} \qquad
\chi(B) = -\frac{1}{L^2}\frac{\partial^2 \Omega}{\partial
B^2}.
\ee
In turn, the grand canonical potential $\Omega$ is directly computed from
the ground state HF energy by $\Omega = E^{\rm HF}_g - TS - N \mu$, where
$S$ is the entropy and $\mu$ the chemical potential, both functions of
$B$.

In the remaining of the paper, the magnetization is given in units
of the effective Bohr magneton $\mu^*_B=e \hbar/2m^* c$ and the
magnetic susceptibility is expressed in units of the Landau
susceptibility, namely $|\chi_L|=e^2/(12 \pi m^{*} c^2)$. For
simplicity, the non-interacting case is referred to as
``$L/a^{*}_{B}=0$".

\subsection{Exchange contribution}
\label{subsec:direct_exchange}

Following Eq. (\ref{eq:EHFdireto}), one can distinguish different
contributions to the magnetic susceptibility, namely
\be
\chi = \chi^{\rm kin} + \chi^{\rm d} + \chi^{\rm x}
\ee
arising from the kinetic, direct, and exchange interaction terms of
the HF ground state energy $E^{\rm HF}_g(B)$. As shown bellow, it is
instructive to compare the later with the susceptibility computed for
non-interacting electrons in a square cavity, $\chi^{\rm non-int}$.

\begin{figure}
\includegraphics[width=8cm]{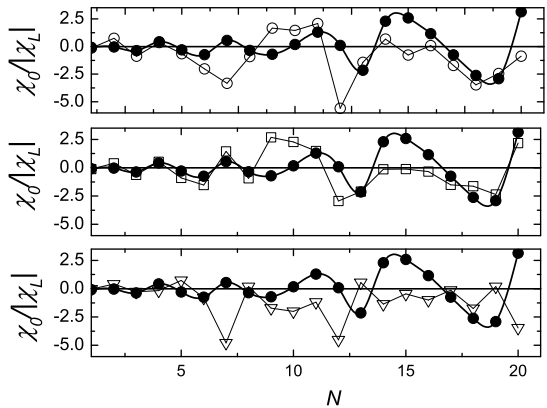}
\caption{Top panel: zero-field susceptibility $\chi_0$ as a
function of the particle number $N$ for the non-interacting (thick
solid line with filled circles) and $r_s=1.5$ (open circles) cases
at $k_BT = \Delta/2$. Middle panel: the same for the kinetic term
(open squares). Bottom panel: the same for the exchange
contribution (open triangles) to the magnetic susceptibility.}
\label{fig:chi0_N}
\end{figure}

The zero-field susceptibility $\chi_0(T)\equiv\chi(B= 0,T)$ is
obtained for different $N$ up to $N=20$ at $k_B T = \Delta/2$.
The results are shown on Fig.~\ref{fig:chi0_N}. The noninteracting
susceptibility $\chi^{\rm non-int}$ (filled circles) oscillates
with increasing amplitude as $N$ is varied (or equivalently, as a
function of $k_F L$). This is in agreement with the previous
semiclassical
calculations.\cite{Prado94,vonOppen94,Agam94,Ullmo95} As we
include the electron-electron interaction included, keeping
$r_s\approx1.5$ for all $N$, the interacting susceptibility (open
circles) sensibly deviates from the noninteracting case. We find
that such deviations are mainly caused by the exchange
contribution to the susceptibility. While the kinetic contribution
(squares) resembles the noninteracting situation, the exchange
term (triangles) exhibits rather large fluctuations, which are the
main responsible for the deviations in the total interacting
susceptibility.

\begin{figure}
\includegraphics[width=8cm]{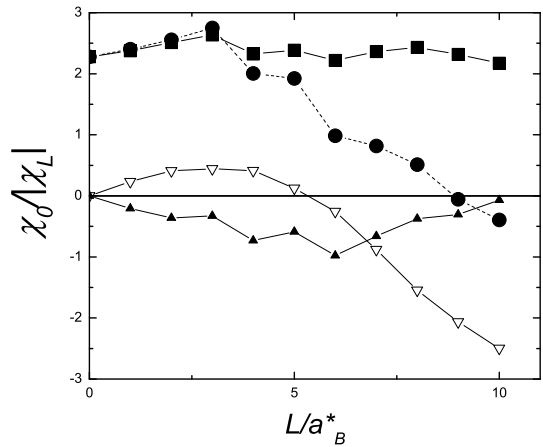}
\caption{Zero-field susceptibility $\chi_0$ (circles) as a
function of interacting strength $L/a^*_B$ for $N=20$ electrons
and $k_B T = \Delta/2$. Kinetic (squares), direct interaction
(filled triangles), and exchange interaction (empty triangles)
contributions to $\chi_0$ are shown for comparison.}
\label{fig:Chi0vsLa_N20}
\end{figure}

The exchange contribution to the susceptibility increases with the
interaction strength $L/a^*_B$ and is already of the order of the
kinetic contribution for $r_s \approx 1.5$. This is illustrated in
Fig.\ \ref{fig:Chi0vsLa_N20}, where we chose a maximum of the
noninteracting $\chi_0^{\rm non-int}(N)$, namely, $N=20$ and vary
$L/a^*_B$.
For low values of $L/a^*_B$, the paramagnetic kinetic term $\chi^{\rm kin}$
(filled squares) dominates and stays almost constant while the
total susceptibility decreases. This behavior of the total
susceptibility is dictated by the exchange term $\chi^{\rm x}$, which is
initially of the order of the direct contribution. As the
interaction strength increases, both the absolute value of the
direct and exchange terms increase. It is interesting to notice
that $\chi^{\rm x}$ overcomes $\chi^{\rm d}$ for $L/a^*_B \approx 5$
$(r_s \approx 0.63)$ and is of the same order as the kinetic term
for $L/a^*_B \approx 10$ $(r_s \approx 1.26)$.

This is a somewhat unexpected result, since the direct
contribution $V^{\HF}_d$ to the ground state energy  is 3 to 4
times {\it larger} than the exchange contribution $V^{\HF}_x$ in
this range of $r_s$. It turns out, however, that the magnetic
$V^{\HF}_x$ is more sensitive to variations of the magnetic filed than
$V^{\HF}_d$, giving rise to the larger exchange contribution.

\subsection{Magnetic field effects}
\label{subsec:magfield}

\begin{figure}
\includegraphics[width=8cm]{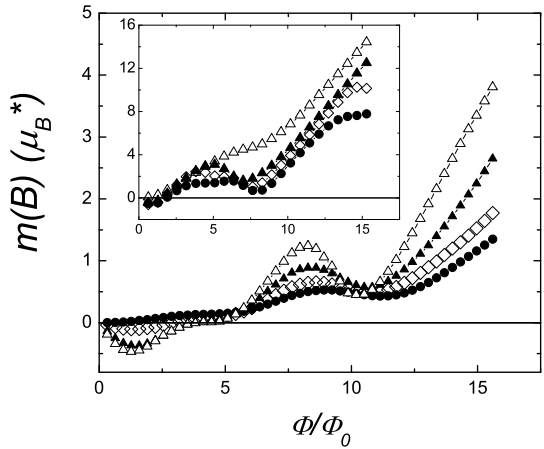}
\caption{Magnetization $m(B)$ as a function of $B = \Phi/L^2$ for
$L/a^*_B=0$ (circles), $2$ (diamonds), $5$ (filled triangles), and
$8$ (open triangles). Here $N=10$ and $k_BT=\Delta$. Inset: $N=20$
for $L/a^*_B=0$ (circles) $3$ (diamonds), $5$ (filled triangles),
and $10$ (open triangles) at $k_BT=\Delta/2$. }
\label{fig:magB}
\end{figure}

\begin{figure}
\includegraphics[width=8cm]{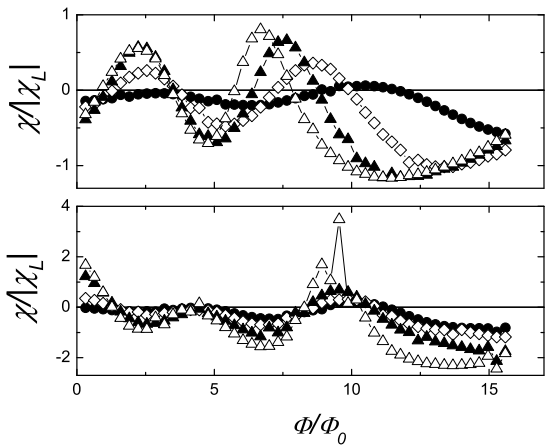}
\caption{Magnetic susceptibility $\chi/|\chi_L|$ as a function of
the magnetic flux for $N=5$ (top) and $N=10$ (bottom) with
$L/a^*_B=0$ (circles), $2$ (diamonds), $4.7$ (filled triangles),
and $8$ (open triangles). The temperature is $k_BT=\Delta$.
 }
\label{fig:susB}
\end{figure}

The electron-electron interaction also induces some nontrivial
effects in the magnetic field dependence of both the magnetization
and the magnetic susceptibility. The results for $m(B)$ and
$\chi(B)$ in the Coulomb ($\kappa=0$) case are shown in
Figs.\ \ref{fig:magB} and \ref{fig:susB} for different $N$
and $L/a^*_B$. In this subsection we take  $k_B T$ of the order of
$\Delta$, to wash out effects due to level crossings, while preserving
the quantum effects due to long energy range spectral correlations.

In the noninteracting case, the magnetization curves for different
$N$ values are very similar. For low fields, small oscillations arise
and, as $\Phi$ increases, a positive magnetization  phase appears.
When the interaction between the electrons is included, the
magnetization curves for $N=10$ and $N=20$ are quite different
(see Fig. \ref{fig:magB}). In particular, for $L/a^{*}_{B} = 10$,
the orbital magnetization is about four times larger for $N=20$
then for  $N=10$. This suggests that large systems can
display strong orbital magnetization effects, in line with the
experimental results. \cite{Grundler02}

For $N=5$, a clear oscillatory behavior is seen on $\chi(B)$ (see
top of Fig. \ref{fig:susB}). The amplitudes are of order
$|\chi_L|$ and, as the interaction strength is increased, both the
amplitude and frequency also increase. This behavior is similar to
the de Hass-von Alphen effect observed in metals. In that case,
both the amplitude and frequency of $\chi(B)$ oscillations are
proportional to the chemical potential of the system. In the
single-particle effective potential approximation, an increase in
the interaction strength is equivalent to an increase in the
effective chemical potential of the HF levels, which corroborates
the analogy.

\begin{figure}
\includegraphics[width=9cm]{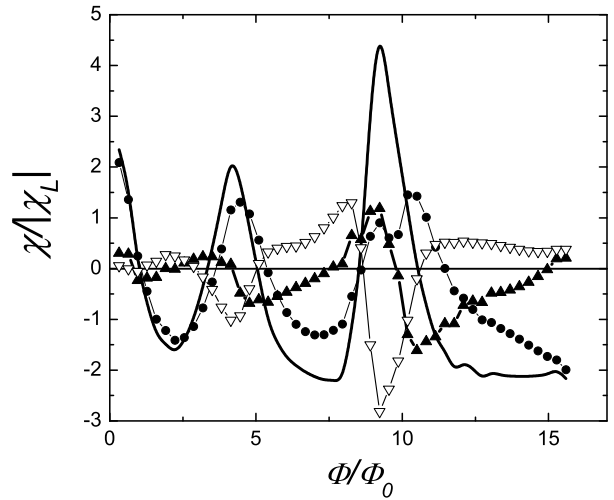}
\caption{Magnetic susceptibility $\chi/\chi_L$ (thick solid line)
as a function of $\Phi/\Phi_0$ for $N=10$ at $k_BT = \Delta/2$ and
$L/a^*_B=5.0$. The symbols correspond to the kinetic (circles),
direct (filled triangles) and exchange contributions (open
triangles). }
\label{fig:Chi_dx_N10loa5_vs_B}
\end{figure}

The oscillations are modified with increasing number of particles,
see bottom panel of Fig.\ \ref{fig:susB}. For larger values of
$L/a^{*}_{B}$, $\chi(B)$ displays fluctuations near
$\Phi/\Phi_0=10$ due to crossings between the Hartree-Fock
single-particle states at the Fermi level, as discussed in Sec.\
\ref{sec:singleparticleproperties}. Notice that although the level
crossing jumps in the single-particle spectrum seem to have
already disappeared at $k_BT = \Delta$, see Fig.\ \ref{fig:EspHF},
they are greatly enhanced in $\chi(B)$, due the second derivative
in Eq.\ (\ref{eq:defchi}).

We observe that the exchange part plays an important role to
for the computation of the $B$ dependence on the magnetic
susceptibility $\chi$.
In Fig.\ \ref{fig:Chi_dx_N10loa5_vs_B} all contributions are shown as a
function of $\Phi/\Phi_0$ for a lower temperature, namely $k_BT =
\Delta/2$ and $r_s=0.9$. The pronounced paramagnetic peak around
$\Phi/\Phi_0 \approx 10$ (also observed at the bottom panel of
Fig. \ref{fig:susB}) is mainly due to the exchange contribution
$\chi^{\rm x}$. Notice that frequently $\chi^{\rm d}$ and
$\chi^{\rm x}$ give contributions of opposite signs appears as
$\Phi/\Phi_0$ is varied.

It is worth mentioning that these results are somewhat different
from the ones obtained by the exact diagonalization for the
two-electron case, \cite{Gregorio02} where the electron spin plays
an important role on the orbital properties of the system. In that
situation, singlet-triplet transitions give rise to
low-temperature peaks of about $3|\chi_L|$ of magnitude in the
orbital susceptibility but no oscillatory behavior is found on
$\chi(B)$. Also, the magnetization $m(B)$ in the exact $N=2$ case
is always negative, even in the large field regime.

\begin{figure}
\includegraphics[width=8cm]{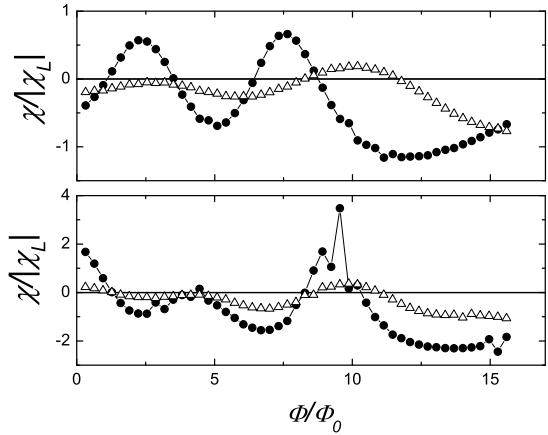}
\caption{ Magnetic susceptibility $\chi(B)$ for different
interaction ranges (Coulomb [circles] and $\kappa^{-1}=L/10$
[triangles]) for $N=5$ (top) and $N=10$ (bottom). }
\label{fig:CoulXa10}
\end{figure}

In Fig. \ref{fig:CoulXa10}, we compare the susceptibilities in the
case of Coulomb and short-range potentials. In order to get a
qualitative comparison, we consider the strong-screening regime,
using $\kappa^{-1}=L/10$.

In this short-range limit, the interaction between the electrons
is exponentially suppressed for distances larger than the
screening length $\kappa^{-1}$ and the overall magnetic response,
which is averaged over the whole dot, is then similar to the
noninteracting case. This can be clearly seen in Fig.
\ref{fig:CoulXa10}where the curves for $\kappa^{-1}=L/10$ are
quite similar to those for $L/a^{*}_{B}=0$ depicted in Fig.
\ref{fig:susB}, showing that the shielding of the
electron-electron interaction is quite effective and a
noninteracting picture is a good approximation for the
thermodynamic properties of the system.

For such low screening lengths, the Fermi wavelength $\lambda_F
\sim \pi L/\sqrt{N}$ is larger than $\kappa^{-1}$ by one order of
magnitude. In the regime $\lambda_F \gg \kappa^{-1}$, the 2D
screened interaction can be well approximated by a $\delta$-type
contact interaction. \cite{Ullmo98} Our results show that the
noninteracting picture captures the main features of such
approximation in the Hartree-Fock scheme. Therefore, one has to
consider larger screening lengths in order to see residual
interaction effects in the magnetization and magnetic
susceptibility.

\subsection{Temperature dependence}
\label{subsec:temp}

\begin{figure}
\includegraphics[width=8cm]{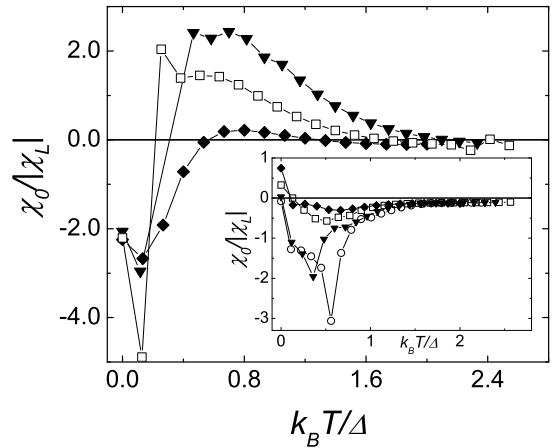}
\caption{ Temperature dependence of the zero-field susceptibility
$\chi_0(T)$ for $N=10$ and $L/a^*_B=0$ (diamonds), $2$ (squares)
and $5$ (triangles). In the inset, the same is shown for $N=5$
with an additional curve for $L/a^*_B=8$ (circles).
}
\label{fig:Sus0_T}
\end{figure}

The zero-field susceptibility decays with temperature, as expected, but
the decay rate is rather large. For $k_BT/\Delta = 3$, $\chi_0$ is already
negligible compared to the $T=0$ value
(see Fig. \ref{fig:Sus0_T}).
In the intermediate range of $k_BT  \sim \Delta$, non-universal features
arise.
As an example, the inset of Fig.\ \ref{fig:Sus0_T} shows our result for a
dot with $N=5$ electrons: $\chi_0(T)$ is negative (except for $T \sim 0$)
and has pronounced minima at $k_BT = \Delta/2$, even in the non-interacting
case.
For $N=10$, $\chi_0(T)$ displays both positive (for $k_BT<0.4\Delta$ and
$k_BT>\Delta$) and negative values ($0.4 \Delta<k_BT<\Delta$), indicating a
diamagnetic-paramagnetic transition. In both cases,
the interaction-induced susceptibility, $\chi-\chi^{\rm non-int}$, is
positive, indicating a paramagnetic contribution due to the electron-electron
interaction.

Such a diamagnetic-paramagnetic transition for increasing $N$ is
in agreement with previous theoretical results. For $N=2$, exact
diagonalization results \cite{Gregorio02} show that $\chi_0(T)$ is
diamagnetic for a temperature range up to $10 \Delta$. On the
other hand, semiclassical analysis
\cite{Prado94,vonOppen94,Ullmo95} suggests a paramagnetic
$\chi_0(T)$ for $N \sim 100$ . Therefore, a
diamagnetic-paramagnetic transition is expected in the
intermediate particle number regime.

\section{Concluding remarks}
\label{sec:conclusions}

We have considered the ground-state properties of a system of $N$
interacting electrons confined in a 2D geometry and subjected to a
perpendicular magnetic field in the finite temperature SCHF
approximation. The magnetic susceptibility was calculated as a
function of the relevant parameters of the system (magnetic field,
number of electrons, temperature, and strength and range of the
particle interaction). The ground-state energy was obtained for
both the Coulomb interaction and the short-range Yukawa potential.

Our results show that the electron-electron interaction introduces
nontrivial effects in the magnetic properties of the system. The
magnetic susceptibility shows de Hass-van Alphen-like oscillations
which are enhanced as the interaction strength increases. The
magnetization increases when more electrons are added in the dot,
which indicates that a strong orbital magnetization should be
expected for larger systems. For a higher number of electrons, new
features arise, including strong diamagnetic fluctuations as a
function of the magnetic field.

The zero-field susceptibility $\chi_0(T)$ shows both paramagnetic
and diamagnetic phases as a function of the temperature. We found
that $\chi_0(T) \rightarrow 0$ as $T$ increases and the
susceptibility induced by interaction is positive, yielding a
paramagnetic contribution to $\chi_0$ irrespective of the value of
$N$. However, non-universal $N$-dependent features appear in the
intermediate temperature range of $k_BT \approx \Delta$.

\acknowledgments This work was supported by Brazilian funding
agencies FAPESP, CNPq, PRONEX, and by the Instituto do
Mil\^enio de Nanoci\^encias. CHL thanks Centro Brasileiro de
Pesquisas F\'{\i}sicas for the hospitality. We thank M.~A.~M.~de
Aguiar, F.~Toscano, and R.~O.~Vallejos for helpful discussions.

\appendix
\section{Hartree-Fock equations in matrix form}

We solve the HF self-consistent equations (\ref{eq:HFeq})
iteratively by diagonalization.
The eigenfunctions of the square billiard of side length
$L$, namely
\be
\label{eq:fi2d}
\varphi_{\alpha}(x,y) = \frac{2}{L}\sin\left(\frac{ m \pi }{L} x \right)
                                \sin\left(\frac{ n \pi }{L} y \right)\,.
\ee
where $\alpha\equiv(m,n)$, separated into the square four point symmetry
classes, form the basis set $\{\psi_\alpha\}$.
The later are also eigenbasis of the operator $R_{\pi/4}$ that rotates the
coordinates by $\pi/4$. The eigenvalues $R_{\pi/4}$, namely $c_\alpha=
+1,-1,+i$ and $-i$, label the square billiard symmetry classes.
More details can be found in Ref. \onlinecite{Gregorio02}.

The bottleneck for HF method is the calculation of the two-body
electron interaction integrals, namely
\be I_{\alpha \gamma \beta \delta}
\equiv  \int \!\!d{\bf r} \!\int\!\! d{\bf r}^\prime \,\psi^{c_\alpha
*}_{\alpha}({\bf r}) \psi^{c_\gamma *}_{\gamma}({\bf r}^\prime)
v({\bf r},{\bf r}^\prime) \psi^{c_\beta}_{\beta}({\bf r})
\psi^{c_\delta}_{\delta}({\bf r}^\prime). \ee
Here the advantage of
separating the basis into symmetry classes comes into play. Many
of the matrix elements $I_{\alpha \gamma \beta \delta}$ are zero,
depending on the single-particle symmetry classes involved.
The ``selection rule" can be summarized as follows: given the
single-particle
symmetry class of the states $\psi^{c_{\alpha}}_{\alpha},
\psi^{c_{\beta}}_{\beta}, \psi^{c_{\gamma}}_{\gamma},$ and $
\psi^{c_{\delta}}_{\delta}$, the matrix element $I_{\alpha \gamma
\beta \delta}$ will be only nonzero  if \cite{Gregorio02}
\ba
c_{\alpha} \otimes c_{\beta} = c_{\gamma} \otimes c_{\delta}\,.
\ea
Therefore, only states with the same two-particle symmetry class are
coupled by the interaction potential.

Let us now express the HF equations in terms of the two-body residual
interaction integrals $I$.
The HF orbital wave functions read
\begin{equation}
\phi_i({\bf r}) = \sum_{\alpha=1}^M C_{i \alpha}
\psi^{c_{\alpha}}_{\alpha}({\bf r}) \,.
\end{equation}
We typically truncate the basis set taking at least the $M = 50$
lowest square billiard energy states. The roman labels refer to
the HF orbitals $\phi$, whereas the greek ones to the basis set $\psi$.
The resulting matrix form of
Eq.\ (\ref{eq:HFeq}) is
\be
\sum_{\beta} (h_{\alpha \beta} + v^{\HF}_{\alpha \beta})
C_{i \beta } = \varepsilon^{\HF}_i C_{i \alpha}\,.
\ee
Here $v^{\HF}_{\alpha \beta}  =  v^{\rm d}_{\alpha \beta} -
v^{\rm x}_{\alpha \beta}$, that are given by
\ba
v^{\rm d}_{\alpha \beta} =&& \!\!\!\!\!\!\int\!\! d{\bf r}\!
\int\!\! d{\bf r}^\prime \,\rho({\bf r}^\prime, {\bf r}^\prime)
      \psi^{c_{\alpha}*}_{\alpha}({\bf r})v({\bf r},{\bf r}^\prime)
      \psi^{c_{\beta}  }_{\beta }({\bf r}) \nonumber \\
v^{\rm x}_{\alpha \beta} =&& \!\!\!\!\!\!\int \!\!d{\bf r}\!
\int \!\!d{\bf r}^\prime \, \rho({\bf r}, {\bf r}^\prime)
      \psi^{c_{\alpha}*}_{\alpha}({\bf r})v({\bf r},{\bf r}^\prime)
      \psi^{c_{\beta}}_{\beta}({\bf r'}),
\ea
with the density matrix $\rho$ given by
\be
\rho({\bf r}, {\bf r}^\prime) = \sum_{i=1}^M n_i^{{}}\, \phi_i^{{}}
 ({\bf r})\phi_i^*({\bf r}^\prime),
\ee
where $n_i$ is the Fermi occupation number of the $i$th HF orbital.
Notice that since $\rho$ does not distinguish different symmetry
classes, the HF mean field can effectively mix them.
Hence, the HF potential reads
\be
v^{\rm HF}_{\alpha \beta}= v^{\rm d}_{\alpha \beta} -
v^{\rm x}_{\alpha \beta} = \sum_{\gamma=1}^M \sum_{\delta=1}^M
D_{\gamma\delta} \Big( I_{\alpha \gamma \beta \delta} -
I_{\alpha \gamma \delta\beta}\Big)\,,
\label{eq:vsmart}
\ee
where, by introducing
\be
 D_{\gamma \delta}  \equiv  \sum_{i=1}^{M}  n_i C^{*}_{i \gamma}
C^{}_{i \delta}\,.
\ee
we eliminate one sum over the single-particle orbitals.
The computation of (\ref{eq:vsmart}) is speed up by exploring the sparce
nature of $I_{\alpha \gamma \delta\beta}$.

Notice that the HF potential matrix elements will be, in the
general case, complex numbers since the basis itself is complex.
The remaining of the numerical implementation is very standard.
The convergence of the ground-state energies is obtained at
iteration $n$ if
$|E^{\HF}_{n+1}-E^{\HF}_{n}|/E^{\HF}_{n}<10^{-5}$.

%



\begin{thebibliography}{99}

\bibitem{Mesoreview}
   L.~L. Sohn, L.~P. Kowenhoven and G. Schon, eds.,
        {\it Mesoscopic Electron Transport} (Kluwer, New York, 1997).

\bibitem{Aleiner02}
   I.~L. Aleiner, P.~W. Brouwer, and L.~I. Glazman,
        Phys. Rep. {\bf 358}, 309 (2002).

\bibitem{Ullmo01}
   D. Ullmo and H.~U. Baranger,
         Phys. Rev. B {\bf 64}, 245324 (2001).

\bibitem{Reimann02}
   S.~M. Reimann and M. Manninen,
        Rev. Mod. Phys. {\bf 74}, 1283 (2002).

\bibitem{Levy93}
   L.~P. L\'{e}vy, D. H. Reich, L. Pfeiffer, and K. West,
         Physica B {\bf 189}, 204 (1993).

\bibitem{Prado94}
   S.~D. Prado and M.~A.~M. de Aguiar,J.P. Keating and R.Egydio de Carvalho
         J. Phys. A {\bf 27}, 6091 (1994).

\bibitem{vonOppen94}
   F. von Oppen,
        Phys. Rev. B {\bf 50}, 17151 (1994).

\bibitem{Agam94}
   O. Agam,
        J. Phys. I {\bf 4}, 697 (1994).

\bibitem{Ullmo95}
   D. Ullmo, K. Richter, and R.~A. Jalabert,
        Phys. Rev. Lett. {\bf 74}, 383 (1995);
   K.~Richter, D.~Ullmo, and R.~A.~Jalabert,
        Phys. Rep. {\bf 276}, 1 (1996).

\bibitem{Gurevich97}
   E. Gurevich and B. Shapiro,
        J. Phys. I {\bf 7}, 807 (1997).

\bibitem{Terra98}
   M.~O.~Terra, M.~L.~Tiago, M.~A.~M.~de Aguiar,
        Phys. Rev. E {\bf 58} 5146 (1998).

\bibitem{Richter96b}
   K. Richter, D. Ullmo, and R.~A. Jalabert,
        Phys. Rev. B {\bf 54}, R 5219 (1996);
        J. Math. Phys. {\bf 37}, 5087 (1996).

\bibitem{Ullmo98}
   D. Ullmo, H. U. Baranger, K. Richter, F. von Oppen, and R. A. Jalabert,
        Phy. Rev. Lett. {\bf 80}, 895 (1998).

\bibitem{Grundler02}
   M.~P. Schwarz, D. Grundler, M. Wilde, Ch. Heyn, and D. Heitman,
        J. Appl. Phys. {\bf 91}, 6875 (2002).

\bibitem{Maksym92}
    P.~A. Maksym and T. Chakraborty,
        Phys. Rev. B {\bf 45}, 1947 (1992).

\bibitem{Wagner92}
    M. Wagner, U. Merkt, and A.~V. Chaplik,
        Phys. Rev. B {\bf 45}, 1951 (1992).

\bibitem{Creffield00}
   C.~E. Creffield, J.~H. Jefferson, S. Sarkar, and D.~L.~J. Tipton,
        Phys. Rev. B {\bf 62}, 7249 (2000).

\bibitem{Gregorio02}
   L.~G.~G.~V. Dias da Silva and M.~A.~M. de Aguiar,
        Phys. Rev. B {\bf 66}, 165309 (2002).

\bibitem{Fogler94}
    M.~M. Fogler, E.~I. Levin, and B.~I. Shklovskii,
        Phys. Rev. B {\bf 49}, 13767 (1994).

\bibitem{Ahn99}
   K.-H. Ahn, K. Richter, and I.-H. Lee,
         Phys. Rev. Lett. {\bf 83}, 4144 (1999).


\bibitem{Ando82}
    T. Ando, A.~B. Fowler, and F. Stern,
        Rev. Mod. Phys. {\bf 54}, 437 (1982).

\bibitem{Candido98}
    L. Candido, J.-P. Rino, N. Studart, and F. M. Peeters,
        J. Phys.: Conden. Matter {\bf 10}, 11627 (1998).

\bibitem{Tamura97}
   H. Tamura and M. Ueda,
         Phys. Rev. Lett. {\bf 79}, 1345 (1997).

\bibitem{Dean01}
   D.~J. Dean, M.~R. Strayer, and J.~C. Wells,
         Phys. Rev. B {\bf 64}, 125305 (2001).

\bibitem{Pfannkuche93}
   D. Pfannkuche, V. Gudmundsson, and P.~A. Maksym,
        Phys. Rev. B {\bf 47}, 2244 (1993).

\bibitem{Vallejos02}
   R.~O. Vallejos, C.~H. Lewenkopf, and Y. Gefen,
         Phys. Rev. B {\bf 65} 085309 (2002).

\bibitem{Gefen02}
   Y. Gefen, R. Berkovits, I.~V. Lerner, and B.~L. Altshuler,
         Phys. Rev. B {\bf 65}, 081106 (2002).


\end{thebibliography}
\end{document}